\pdfoutput=1
\documentclass[aps,prb,reprint,amsmath,amssymb,floatfix]{revtex4-2}
\usepackage{graphicx}
\usepackage{bm}
\usepackage{dcolumn}
\usepackage[colorlinks=true,allcolors=blue]{hyperref}

\newcommand{\M}{m^*/m}
\newcommand{\OO}{\mathcal{O}}
\newcommand{\ALP}{A_{\mathrm{LP}}}

\begin{document}

\title{Fourth-order perturbation theory for the Fr\"ohlich polaron:\\
Analytic structure of the weak-coupling series and the crossover to strong coupling}

\author{Sankar Das Sarma}
\affiliation{Condensed Matter Theory Center and Joint Quantum Institute, Department of Physics,
University of Maryland, College Park, Maryland 20742, USA}

\date{\today}

\begin{abstract}
We calculate the Rayleigh-Schr\"odinger (RS) perturbation series for the
ground-state energy and the effective mass of the three-dimensional
Fr\"ohlich polaron through fourth order in the coupling constant $\alpha$,
obtaining $\M=1+\alpha/6+0.02362763\,\alpha^2+0.0031540\,\alpha^3
+4.080(8)\times10^{-4}\,\alpha^4+\cdots$, where the third- and fourth-order
coefficients are new. Because the Fr\"ohlich interaction is a form-bounded
perturbation of an isolated nondegenerate ground state at fixed total
momentum, the series has a nonzero radius of convergence, and we analyze it
with the ratio and Pad\'e methods appropriate to convergent series. The
coefficients are strikingly regular: they are described by
$\M\simeq(1-\alpha/R)^{-s}$ with $R=8.55(10)$ and $s=1.42(5)$, and with
$R$ and $s$ fixed by the first two orders this form predicts the
third-order coefficient of the inverse mass to $0.5\%$ and the
fourth-order one within its uncertainty. The couplings at which successive RS truncations of the inverse
mass vanish, $6\to7.34\to7.76\to7.96(4)$, the first of which is the
textbook ``breakdown'' value $\alpha=6$, are slowly converging estimates of
$R$. Pad\'e approximants built from the weak-coupling coefficients alone
agree with diagrammatic Monte Carlo results for the mass to better than
$1\%$ for $\alpha\le5$ and to $5\%$ at $\alpha=6$, and for the energy within
the combined uncertainties up to $\alpha=8$. Combining the
Vivanti-Pringsheim theorem with the known analyticity of the polaron ground
state for all real $\alpha$ and with the rigorous Pekar and Landau-Pekar
strong-coupling asymptotics, we show that neither series can have
coefficients of ultimately constant sign, although every coefficient known
at present has the same sign. The apparent singularity at $\alpha\simeq8.5$
must therefore be a complex-conjugate pair close to the positive real axis.
It marks the crossover to strong coupling: the weak-coupling series locates
this crossover accurately, but it cannot be continued through it without
information that, by our estimate, first appears in the coefficients at
orders of several tens.
\end{abstract}

\maketitle

\section{Introduction}
\label{sec:intro}

The Fr\"ohlich polaron, a single band electron coupled to dispersionless
longitudinal-optical phonons through the long-range polar interaction
\cite{Frohlich1954,LeeLowPines1953}, is one of the oldest nontrivial
problems of quantum field theory in condensed matter and the standard
testing ground for nonperturbative methods
\cite{Feynman1955,DevreeseAlexandrov2009}. It has a single dimensionless
coupling constant $\alpha$ and no exact solution. Weak coupling is
controlled by Rayleigh-Schr\"odinger (RS) perturbation theory. Strong
coupling is controlled by the adiabatic theory of Pekar
\cite{Pekar1954,LandauPekar1948,Miyake1975,Miyake1976}, whose leading
predictions are now theorems: the ground-state energy obeys
$E_0=e_{\rm P}\alpha^2+o(\alpha^2)$ with $e_{\rm P}=-0.108513$
\cite{DonskerVaradhan1983,LiebThomas1997,BrooksSeiringer2023}, and the
effective mass obeys
\begin{equation}
 \lim_{\alpha\to\infty}\alpha^{-4}\,\M=\ALP\simeq0.0227019 ,
 \label{eq:LP}
\end{equation}
the sharp upper bound having been proved by Brooks and Seiringer
\cite{BrooksSeiringer2024}, lower bounds of the correct order by Sellke
and by Bazaes \emph{et al.}\ \cite{Sellke2024,BMSV2023}, and the sharp
lower bound by Brooks \cite{Brooks2026}. In between, the numerically
exact diagrammatic Monte Carlo (DMC) method
\cite{ProkofevSvistunov1998,Mishchenko2000,Hahn2018} shows a rapid but
smooth growth of the mass, in agreement with the theorem of Gerlach and
L\"owen that the ground state is analytic in $\alpha$ on the whole positive
real axis, so that no ``self-trapping'' phase transition exists
\cite{GerlachLowen1991}.

This paper asks a simple question about this well-characterized problem:
what do the first few weak-coupling coefficients know about the crossover
to strong coupling? The question has a familiar folklore answer. The
leading RS dispersion gives $m/m^*=1-\alpha/6$, which vanishes at
$\alpha=6$, and this value is still widely quoted as the coupling at which
perturbation theory breaks down and the polaron begins to self-trap
\cite{DevreeseAlexandrov2009,deMelo2023}. Whether that number survives
higher orders, and what it means if it does not, has not been examined,
because the mass series has been known only through $\OO(\alpha^2)$
\cite{HohlerMullensiefen1959,Roseler1968}. The energy series is known
further: the $\alpha^3$ coefficient was obtained by Smondyrev
\cite{Smondyrev1986,SelyuginSmondyrev1989}, and the $\alpha^4$ and
$\alpha^5$ coefficients by Rosenfelder with a cumulant method that does not
yield the mass \cite{Rosenfelder2009}. Higher-order contributions to the
mass have been discussed within variational and path-integral treatments
\cite{Larsen1968,KholodenkoFreed1983,Larsen1984}, but these are not RS
coefficients.

Here we compute the RS coefficients of the effective mass at third and
fourth order and use the resulting four-term series, together with the
five-term energy series, to address the question above. Our main results
are as follows.

(i) The mass series through fourth order is
\begin{align}
 \frac{m^*}{m}={}&1+\frac{\alpha}{6}+0.02362763\,\alpha^2
 +0.0031540\,\alpha^3\nonumber\\
 &+4.080(8)\times10^{-4}\,\alpha^4+\OO(\alpha^5) ,
 \label{eq:massseries}
\end{align}
and the fourth-order energy coefficient obtained in the same calculation,
$-5.31(10)\times10^{-5}$, agrees with Rosenfelder's
$-5.33(1)\times10^{-5}$.

(ii) The RS series for the energy and the mass have a nonzero radius of
convergence. This is a direct consequence of analytic perturbation theory
and contrasts with the factorially divergent series of quantum
electrodynamics (QED) and other field theories
\cite{Dyson1952,Beneke1999}. The continuation problem is therefore not one
of Borel resummation or resurgence, but of analytic continuation of a
convergent series beyond its circle of convergence, exactly as in the
analysis of high-temperature series in critical phenomena
\cite{DombSykes1957,Guttmann1989,BakerGravesMorris1996}.

(iii) Analyzed in this way, the coefficients are remarkably regular. They
are described by the two-parameter form $\M\simeq(1-\alpha/R)^{-s}$ with
$R=8.55(10)$ and $s=1.42(5)$; when $R$ and $s$ are fixed by the exactly
known first two orders, this form predicts the third-order coefficient of
the inverse mass to $0.5\%$ and the fourth-order one within its
uncertainty.
The zeros
of the successive RS truncations of the inverse mass, $\alpha_N=6$, $7.34$,
$7.76$, $7.96(4)$, converge slowly, as $1/N$, toward $R$; the folklore
value $6$ is simply the first and poorest member of this sequence.

(iv) Pad\'e approximants constructed from the weak-coupling coefficients
alone agree with the DMC mass \cite{Hahn2018} to better than $1\%$ for
$\alpha\le5$ and to $5\%$ at $\alpha=6$, and with the DMC energy within
the combined uncertainties up to $\alpha=8$. Over this range they are
considerably more accurate than Feynman's variational results.

(v) The regularity in (iii) cannot persist. We show, by combining the
Vivanti-Pringsheim theorem with the Gerlach-L\"owen analyticity and the
strong-coupling theorems quoted above, that both the energy and the mass
series must contain infinitely many coefficients of either sign, although
all the coefficients known today have the same sign. The apparent
singularity at $\alpha\simeq8.5$ is therefore a complex-conjugate pair
lying close to, but off, the positive real axis. It is the analytic
fingerprint of the crossover region, where the DMC mass indeed shows its
steepest logarithmic growth. The low-order coefficients fix the modulus of
this singularity but not its distance from the real axis, and simple
estimates indicate that the sign structure required by the exact results
sets in only at orders of several tens.

The weak-coupling series thus knows \emph{where} the crossover to strong
coupling occurs, to about one percent, but not \emph{how} the polaron
passes through it. The contrast with QED is instructive. There, even a
heroic calculation such as the tenth-order anomalous magnetic moment
\cite{Aoyama2012} supplies a short segment of an asymptotic series whose
strong-coupling content is controlled by Borel singularities that few
coefficients cannot locate \cite{Beneke1999,DunneSchubert2002}, and
low-order truncations cannot decide questions such as the existence of a
Landau pole \cite{JianBarnesDasSarma2020}. In the polaron problem the
series converges and its nearest singularity can be located from four
terms, yet the obstruction to reaching strong coupling is of the same
logical kind: the required information is global and resides in the
complex-$\alpha$ plane.

The paper is organized as follows. Section~\ref{sec:formalism} sets up the
RS expansion and its diagrammatic representation.
Section~\ref{sec:coefficients} reports the coefficients.
Section~\ref{sec:analytic} establishes the analytic properties of the
series. Section~\ref{sec:series} contains the series analysis and the
comparison with DMC. Section~\ref{sec:discussion} discusses what the series
does not determine, and Sec.~\ref{sec:conclusions} concludes. Numerical
details are collected in Appendix~\ref{app:numerics}.
Appendix~\ref{app:holstein} reports the corresponding fourth-order
calculation and series analysis for the three-dimensional Holstein polaron.
We consider only three-dimensional models.

\section{Model and perturbation expansion}
\label{sec:formalism}

\subsection{Hamiltonian and Rayleigh-Schr\"odinger series}

We use units $\hbar=\omega_{\rm LO}=m=1$. After the Lee-Low-Pines
transformation \cite{LeeLowPines1953} the Hamiltonian at conserved total
momentum $\bm P$ acts on the phonon Fock space,
\begin{equation}
 H_{\bm P}=\tfrac12(\bm P-\bm P_f)^2+N_f+\sqrt{\alpha}\,V ,
 \label{eq:HP}
\end{equation}
with
\begin{align}
 N_f&=\int_{\bm q}a^\dagger_{\bm q}a_{\bm q},\qquad
 \bm P_f=\int_{\bm q}\bm q\,a^\dagger_{\bm q}a_{\bm q},\nonumber\\
 V&=\int_{\bm q}v_{\bm q}\bigl(a_{\bm q}+a^\dagger_{-\bm q}\bigr),\qquad
 |v_{\bm q}|^2=\frac{2\sqrt2\,\pi}{q^2},
 \label{eq:defs}
\end{align}
and $\int_{\bm q}\equiv\int d^3q/(2\pi)^3$. The unperturbed ground state is
the phonon vacuum $|0\rangle$ with energy $P^2/2$. We write the ground-state
energy as
\begin{align}
 E(\bm P)&=\frac{P^2}{2}+\sum_{n\ge1}\alpha^n\varepsilon_n(\bm P),\nonumber\\
 \varepsilon_n(\bm P)&=e_n+q_nP^2+\OO(P^4),
 \label{eq:EP}
\end{align}
so that
\begin{align}
 E_0&=\sum_{n\ge1}e_n\alpha^n,\qquad
 \frac{m}{m^*}=\sum_{n\ge0}d_n\alpha^n,\nonumber\\
 \frac{m^*}{m}&=\sum_{n\ge0}c_n\alpha^n ,
 \label{eq:series}
\end{align}
with $d_0=c_0=1$ and $d_n=2q_n$ for $n\ge1$. The $q_n$ are the quantities
computed directly; the $c_n$ follow by reversion,
\begin{align}
 c_1&=-d_1,\qquad c_2=d_1^2-d_2,\qquad c_3=-d_1^3+2d_1d_2-d_3,\nonumber\\
 c_4&=d_1^4-3d_1^2d_2+d_2^2+2d_1d_3-d_4 .
 \label{eq:reversion}
\end{align}

Let $Q=1-|0\rangle\langle0|$ and define the reduced resolvent
\begin{equation}
 R_{\bm P}=\frac{Q}{N_f+\tfrac12P_f^2-\bm P\cdot\bm P_f},
 \label{eq:resolvent}
\end{equation}
which is positive for $P<\sqrt2$. Since $V$ changes the phonon number by
one, only even orders in $V$ contribute. With the brackets
\begin{equation}
 B_{r_1r_2\cdots r_{2k-1}}=\langle0|VR^{r_1}VR^{r_2}V\cdots
 R^{r_{2k-1}}V|0\rangle ,
 \label{eq:bracket}
\end{equation}
which are all positive, the RS recursion gives
\begin{align}
 \varepsilon_1={}&-B_1,\qquad \varepsilon_2=B_1B_2-B_{111},\nonumber\\
 \varepsilon_3={}&-B_1^2B_3+B_1B_{112}+B_1B_{121}+B_1B_{211}\nonumber\\
 &-B_1B_2^2+B_{111}B_2-B_{11111},
 \label{eq:e123}
\end{align}
and, at fourth order, the 24-term expression
\begin{align}
 \varepsilon_4={}&B_1^3B_4+3B_1^2B_2B_3+B_1B_2^3\nonumber\\
 &-B_1^2\bigl(B_{113}+B_{122}+B_{131}\nonumber\\
 &\qquad\quad+B_{212}+B_{221}+B_{311}\bigr)\nonumber\\
 &-2B_1B_3B_{111}-2B_1B_2\bigl(B_{112}+B_{121}+B_{211}\bigr)\nonumber\\
 &+B_1\bigl(B_{11112}+B_{11121}+B_{11211}\nonumber\\
 &\qquad\quad+B_{12111}+B_{21111}\bigr)\nonumber\\
 &+B_{111}\bigl(B_{112}+B_{121}+B_{211}\bigr)-B_2^2B_{111}\nonumber\\
 &+B_2B_{11111}-B_{1111111}.
 \label{eq:e4}
\end{align}
We have checked Eqs.~\eqref{eq:e123} and \eqref{eq:e4} against the exact
ground-state energy of finite-dimensional test Hamiltonians having the same
even-odd structure of $V$.

\subsection{Proper histories and the mass curvature}
\label{sec:histories}

\begin{figure*}[t]
 \centering
 \includegraphics[width=0.92\textwidth]{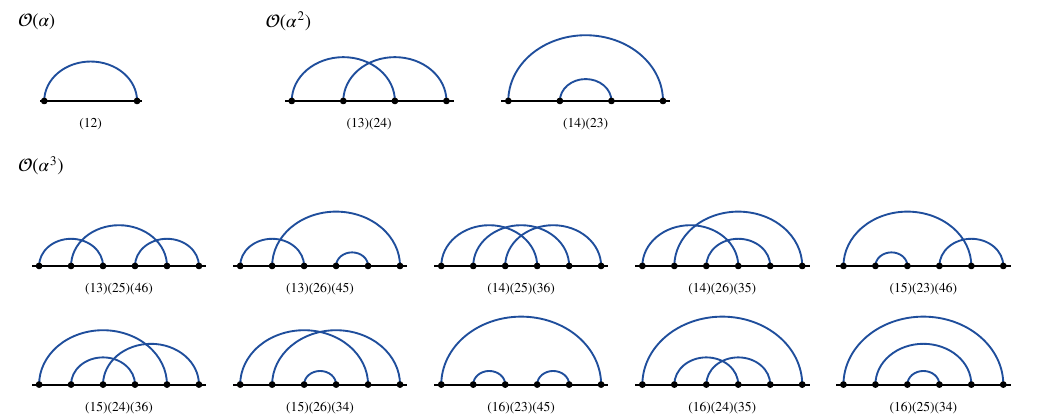}
 \caption{Proper time-ordered phonon contractions contributing to the RS
 brackets of Eq.~\eqref{eq:bracket} through $\OO(\alpha^3)$. The
 horizontal line carries the $2k$ ordered interaction vertices and each
 arc is a phonon; the label lists the contracted vertex pairs. At least
 one phonon is present between any two consecutive vertices. There are
 74 such contractions at $\OO(\alpha^4)$ and 706 at $\OO(\alpha^5)$.}
 \label{fig:diagrams}
\end{figure*}

Each bracket is a sum over time-ordered phonon contractions. A bracket with
$2k$ vertices involves $k$ phonons with momenta $\bm q_1,\dots,\bm q_k$,
labeled in the order of their creation. A \emph{proper history}
$h=(S_1,\dots,S_{2k-1})$ specifies the set $S_j$ of phonons present after
the $j$th vertex; the projector $Q$ requires $S_j\neq\emptyset$ for all
$j$. With
\begin{equation}
 \bm K_j(h)=\sum_{\ell\in S_j}\bm q_\ell,\qquad
 d_j(h)=|S_j|+\tfrac12K_j(h)^2 ,
 \label{eq:Kd}
\end{equation}
one has
\begin{align}
 B_{r_1\cdots r_{2k-1}}={}&\sum_{h\in\mathcal H_k}\int_{\bm q_1\cdots\bm q_k}
 \prod_{\ell=1}^{k}|v_{\bm q_\ell}|^2\nonumber\\
 &\times\prod_{j=1}^{2k-1}\bigl[d_j(h)-\bm P\cdot\bm K_j(h)\bigr]^{-r_j}.
 \label{eq:history}
\end{align}
The number of proper histories is $|\mathcal H_k|=1,2,10,74,706,\dots$ for
$k=1,2,3,\dots$ \cite{OEISA000698}; those for $k\le3$ are shown in
Fig.~\ref{fig:diagrams}. The normalization in Eq.~\eqref{eq:defs} gives
$B_1=1$ at $P=0$.

The total momentum enters only through the denominators in
Eq.~\eqref{eq:history}. Expanding to second order in $\bm P$ and averaging
over its direction gives the coefficient of $P^2$ analytically,
\begin{align}
 \bigl[P^2\bigr]\prod_j(d_j-\bm P\cdot\bm K_j)^{-r_j}={}&
 \frac13\prod_jd_j^{-r_j}\Biggl[\sum_j\frac{r_j(r_j+1)K_j^2}{2d_j^2}
 \nonumber\\
 &+\sum_{\ell<j}\frac{r_\ell r_j\,\bm K_\ell\cdot\bm K_j}{d_\ell d_j}
 \Biggr],
 \label{eq:curvature}
\end{align}
where the products and sums run over all denominators of a given monomial
of Eqs.~\eqref{eq:e123} and \eqref{eq:e4}. No numerical differentiation
with respect to $\bm P$ is needed. At first order
Eqs.~\eqref{eq:history} and \eqref{eq:curvature} give $e_1=-1$ and
$q_1=-1/12$, i.e., the familiar $m/m^*=1-\alpha/6+\OO(\alpha^2)$.

\section{Coefficients through fourth order}
\label{sec:coefficients}

\subsection{Known results}

The second-order coefficients are known in closed form
\cite{HohlerMullensiefen1959,Roseler1968},
\begin{align}
 e_2&=\frac{1}{\sqrt2}-\ln\Bigl(1+\frac{3\sqrt2}{4}\Bigr)=-0.015919622,
 \nonumber\\
 c_2&=\frac43\ln(1+\sqrt2)-\frac23\ln2-\frac{5\sqrt2}{8}+\frac7{36}
 =0.023627630,
 \label{eq:second}
\end{align}
so that $q_2=(1/36-c_2)/2=0.0020750737$. For the energy,
$e_3=-0.000806070$ \cite{Smondyrev1986,SelyuginSmondyrev1989},
$e_4=-5.33(1)\times10^{-5}$, and $e_5=-3.8(2)\times10^{-6}$
\cite{Rosenfelder2009}. We are not aware of previous evaluations of the
RS mass coefficients beyond second order.

\subsection{Third order}

The third-order curvature $q_3$ is the integral over three phonon momenta
of the seven-term integrand that results from inserting
Eqs.~\eqref{eq:history} and \eqref{eq:curvature} into
Eq.~\eqref{eq:e123}. After removal of the three global Euler angles a
six-dimensional integral remains, which we evaluate numerically. The same
code reproduces the exactly known $e_2$, $q_2$, and $e_3$. We obtain
\begin{equation}
 q_3=4.6137\times10^{-5},\qquad
 c_3=2c_1c_2-c_1^3-2q_3=0.0031540 .
 \label{eq:q3}
\end{equation}
The series analysis below is insensitive to changes in the last digit of
$q_3$. As an independent check, $q_3$ was also evaluated with a separately
written code by quasi-Monte Carlo integration of the fully
momentum-symmetrized nine-dimensional integrand, using the exactly known
$e_3$ as a control variate; the result, $4.58(3)\times10^{-5}$, is
consistent with Eq.~\eqref{eq:q3}.

\subsection{Fourth order}

At fourth order there are four phonon momenta, 74 proper histories in the
leading bracket $B_{1111111}$, and 24 monomials in Eq.~\eqref{eq:e4}. The
individual monomials are of order unity, whereas
$\varepsilon_4\sim5\times10^{-5}$, so that the cancellation among them must
take place in the integrand: all monomials are evaluated on the same
four-momentum sample, and the momenta of a product of brackets are simply
the four sampled momenta distributed among its factors. After removal of
the global rotations the integral is nine dimensional. It is evaluated
with scrambled Sobol sequences \cite{Sobol1967,Owen1995}; details and the
raw data are given in Appendix~\ref{app:numerics}. Two independently written
implementations, one of which generates the histories of
Eq.~\eqref{eq:history} recursively, agree pointwise to a relative accuracy
of $4\times10^{-14}$.

\begin{figure}[t]
 \centering
 \includegraphics[width=\columnwidth]{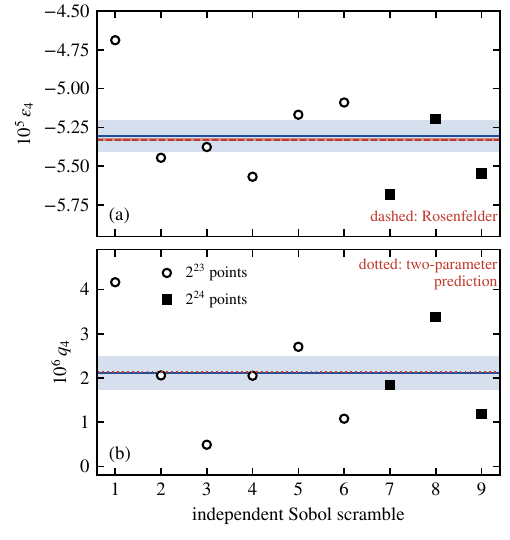}
 \caption{Fourth-order results from nine independent scrambled-Sobol
 integrations. (a) Energy coefficient $e_4$; the dashed line and narrow
 band show the value of Ref.~\cite{Rosenfelder2009}. (b) Mass-curvature
 coefficient $q_4$; the dotted line is the value implied by the
 two-parameter form fixed by $c_1$ and $c_2$ alone, Eq.~\eqref{eq:predq}.
 Solid lines and shaded bands are the means and standard errors of the
 nine runs.}
 \label{fig:fourth}
\end{figure}

The energy and the curvature are accumulated in the same runs. From nine
independent scrambles with $2^{23}$ or $2^{24}$ points each
(Fig.~\ref{fig:fourth}) we obtain
\begin{equation}
 e_4=-5.31(10)\times10^{-5},\qquad q_4=2.1(4)\times10^{-6},
 \label{eq:e4q4}
\end{equation}
where the uncertainties are standard errors over scrambles. The energy
coefficient was not used anywhere in the calculation and agrees with the
ten times more precise value of Ref.~\cite{Rosenfelder2009}; this
validates the fourth-order kernel. Equation~\eqref{eq:reversion} then gives
\begin{equation}
 c_4=4.1222\times10^{-4}-2q_4=4.080(8)\times10^{-4}.
 \label{eq:c4}
\end{equation}
Note that $99\%$ of $c_4$ is fixed by the lower orders through the
reversion; the genuinely new fourth-order information is $q_4$, which is
determined here to about $20\%$. This suffices for the conclusions of this
paper, which rest mainly on the first three orders and are confirmed, not
driven, by the fourth. The coefficients are collected in
Table~\ref{tab:coeffs}.

\begin{table}[t]
\caption{Coefficients of the RS series defined in
Eqs.~\eqref{eq:EP} and \eqref{eq:series}. Entries without uncertainties
are exact to the digits shown. The values of $e_4$ and $e_5$ are from
Ref.~\cite{Rosenfelder2009}; our independent result for $e_4$ is
$-5.31(10)\times10^{-5}$. The entries $q_3$, $c_3$, $q_4$, and $c_4$ are
results of this work.}
\label{tab:coeffs}
\begin{ruledtabular}
\begin{tabular}{llll}
$n$ & \multicolumn{1}{c}{$e_n$} & \multicolumn{1}{c}{$q_n=d_n/2$} &
\multicolumn{1}{c}{$c_n$}\\
\hline
1 & $-1$ & $-1/12$ & $1/6$\\
2 & $-0.015919622$ & $2.0750737\times10^{-3}$ & $0.023627630$\\
3 & $-8.06070\times10^{-4}$ & $4.6137\times10^{-5}$ & $3.1540\times10^{-3}$\\
4 & $-5.33(1)\times10^{-5}$ & $2.1(4)\times10^{-6}$ & $4.080(8)\times10^{-4}$\\
5 & $-3.8(2)\times10^{-6}$ & & \\
\end{tabular}
\end{ruledtabular}
\end{table}

\section{Analytic structure of the series}
\label{sec:analytic}

\subsection{The series converge}
\label{sec:convergence}

The number of contractions grows factorially with the order, as in any
field theory, and it is sometimes assumed that the polaron series must
therefore be asymptotic. It is not. Consider the family
$H_{\bm P}(g)=H^{(0)}_{\bm P}+gV$ with $g=\sqrt\alpha$ complex. Although
$V$ is not operator bounded, because $v_{\bm q}$ is not square integrable
at large $q$, it is infinitesimally form bounded with respect to
$H^{(0)}_{\bm P}=\tfrac12(\bm P-\bm P_f)^2+N_f$
\cite{LiebYamazaki1958,Moller2006}, so that $H_{\bm P}(g)$ is a holomorphic
family in the sense of Kato for all complex $g$
\cite{Kato1976,ReedSimon4}. For $P<\sqrt2$ the unperturbed eigenvalue
$P^2/2$ is nondegenerate and separated by a gap $1-P^2/2$ from the rest of
the spectrum of $H^{(0)}_{\bm P}$, which begins at the one-phonon
threshold. Analytic perturbation theory then guarantees that $E(\bm P;g)$
is analytic in a disk around $g=0$, jointly with its dependence on
$\bm P$. Because the unitary operator $(-1)^{N_f}$ maps $V\to-V$, the
energy is even in $g$ and hence analytic in $\alpha=g^2$ at $\alpha=0$.
The series for $E_0$, $m/m^*$, and $\M$ in Eq.~\eqref{eq:series} therefore
converge for $|\alpha|$ smaller than a radius of convergence that is
strictly positive; we denote by $R$ the radius for the mass. The factorial
growth of the number of histories is no obstacle: convergence is
guaranteed by the operator-theoretic argument and is realized through the
extensive cancellations among the terms of Eqs.~\eqref{eq:e123} and
\eqref{eq:e4}.

This places the polaron in a different class from QED, where the
perturbation series has zero radius of convergence \cite{Dyson1952} and
continuation to strong coupling is a problem of Borel summation and
nonperturbative sectors \cite{Beneke1999,DunneSchubert2002}. For the
polaron, the Borel transform of the series is an entire function and
carries no singularity structure to be exploited. The relevant
mathematical problem is instead the classical one of continuing a
convergent Taylor series beyond its circle of convergence, for which the
natural tools are the ratio method and Pad\'e and Dlog-Pad\'e approximants
\cite{DombSykes1957,Guttmann1989,BakerGravesMorris1996}. The radius $R$ is
the modulus of the singularity of $\M$ nearest to the origin in the
complex-$\alpha$ plane.

\subsection{The coefficients cannot keep a constant sign}
\label{sec:signs}

Two exact results constrain where this singularity can be. First, the
ground-state energy $E(\bm P;\alpha)$ is analytic in $\alpha$ on the whole
real half line $\alpha\ge0$ for small $P$ \cite{GerlachLowen1991}; since
the inverse effective mass is strictly positive for every finite $\alpha$
\cite{Spohn1987,DybalskiSpohn2020,BetzPolzer2022}, the same holds for
$E_0(\alpha)$, $m/m^*$, and $\M$. Second, the strong-coupling asymptotics
$E_0\sim e_{\rm P}\alpha^2$ and Eq.~\eqref{eq:LP} hold. Together with a
classical theorem of function theory these imply the following.

\emph{Proposition.} Each of the coefficient sequences $\{e_n\}$,
$\{c_n\}$, and $\{d_n\}$ contains infinitely many positive and infinitely
many negative terms.

\emph{Proof.} Let $f(\alpha)=\sum_na_n\alpha^n$ stand for $-E_0$ or $\M$,
so that $f$ is real analytic on $[0,\infty)$ and $f\sim C\alpha^k$ for
$\alpha\to\infty$, with $k=2$, $C=|e_{\rm P}|$ or $k=4$, $C=\ALP$.
Suppose that $a_n\ge0$ for all $n\ge n_0$, and let
$g(\alpha)=\sum_{n\ge n_0}a_n\alpha^n$, which differs from $f$ by a
polynomial and has the same radius of convergence $\rho$. If
$\rho<\infty$, the Vivanti-Pringsheim theorem \cite{Titchmarsh1939} states
that a power series with nonnegative coefficients is singular at the
point $\alpha=\rho$ on the positive real axis, which contradicts the
analyticity of $f$ there. If $\rho=\infty$, then for every $n\ge n_0$ and
$\alpha>0$ one has $a_n\alpha^n\le g(\alpha)=\OO(\alpha^{\max(k,n_0-1)})$,
so that $a_n=0$ for all sufficiently large $n$ and $f$ is a polynomial;
the asymptotic law then requires its degree to be $k$ and its leading
coefficient to be $C$. This is false for the energy because $e_3\neq0$,
and for the mass because $c_4\neq\ALP$. The case $a_n\le0$ for all
$n\ge n_0$ is excluded in the same way with $f\to-f$. For the inverse
mass, $f=m/m^*$ is analytic on $[0,\infty)$, tends to zero at infinity,
and equals one at the origin; the same argument applies, since a
polynomial cannot have these properties. $\blacksquare$

The Proposition is to be compared with the coefficients in
Table~\ref{tab:coeffs}: all five known $e_n$ are negative and all four
$c_n$ are positive (the $d_n$ are positive for $n=2,3,4$). The sign
pattern of the known coefficients is thus certainly not the asymptotic
one. Equivalently, the singularities of $E_0$ and $\M$ nearest to the
origin do not lie on the positive real axis. In the generic situation
they form a complex-conjugate pair $\alpha_c=Re^{\pm i\theta}$, and if
$\M\simeq A(1-\alpha/\alpha_c)^{-s}+\mathrm{c.c.}$ near these points,
Darboux's theorem gives for large $n$
\begin{equation}
 c_n\simeq\frac{2|A|}{\Gamma(s)}\,\frac{n^{s-1}}{R^n}\,
 \cos(n\theta-\phi),\qquad \phi=\arg A ,
 \label{eq:darboux}
\end{equation}
so that the coefficients change sign with period $\pi/\theta$ in $n$. The
absence of any sign change at the orders computed so far already
indicates that $\theta$ is small.

\section{Series analysis}
\label{sec:series}

\subsection{Ratio method}

\begin{figure}[t]
 \centering
 \includegraphics[width=\columnwidth]{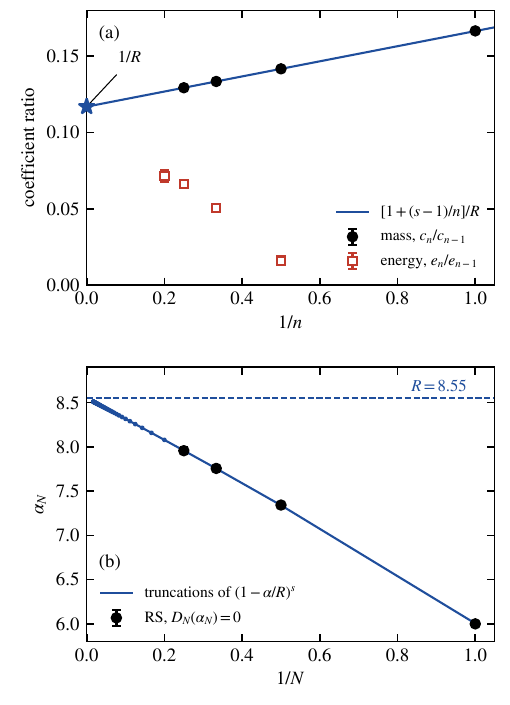}
 \caption{(a) Ratios of successive RS coefficients versus $1/n$ for the
 mass (filled circles) and the energy (open squares). The line is
 Eq.~\eqref{eq:ratio} with $R=8.55$ and $s=1.424$; its intercept (star)
 is $1/R$. (b) Smallest positive zero $\alpha_N$ of the $N$th-order RS
 truncation of the inverse mass, Eq.~\eqref{eq:DN}, versus $1/N$ (filled
 circles), compared with the zeros of the truncated Taylor series of
 $(1-\alpha/R)^s$ for $N\le60$ (line and dots). The first point is the
 textbook value $\alpha=6$.}
 \label{fig:series}
\end{figure}

For a function with a dominant singularity $(1-\alpha/R)^{-s}$ the ratios
of successive coefficients behave as \cite{DombSykes1957,Guttmann1989}
\begin{equation}
 r_n\equiv\frac{c_n}{c_{n-1}}=\frac1R\Bigl[1+\frac{s-1}{n}
 +\OO(n^{-2})\Bigr],
 \label{eq:ratio}
\end{equation}
with the $\OO(n^{-2})$ term absent for a pure power. From
Table~\ref{tab:coeffs},
\begin{equation}
 r_1,\dots,r_4=0.166667,\ 0.141766,\ 0.133487,\ 0.12936(25).
\end{equation}
Plotted against $1/n$ (Fig.~\ref{fig:series}(a)) the four ratios fall on
a straight line to within the size of the symbols. Solving
Eq.~\eqref{eq:ratio} for successive pairs gives the estimates in the upper
part of Table~\ref{tab:estimates}; they agree to one part in $10^3$ where
they are not limited by the uncertainty of $q_4$.

The energy ratios $e_n/e_{n-1}=0.0159$, $0.0506$, $0.0661$, $0.071(4)$
for $n=2,\dots,5$ are also shown in Fig.~\ref{fig:series}(a). They
increase with $n$ and are compatible with a radius of the same magnitude
(unbiased two-point estimates lie between $9$ and $11$), but the energy
series is dominated at low order by its trivial linear term and is not
yet regular enough for a quantitative determination.

\begin{table}[t]
\caption{Estimates of the radius of convergence $R$ and exponent $s$ of
the mass series from the ratio method, Eq.~\eqref{eq:ratio}, and from
Dlog-Pad\'e approximants $[L/M]$ to Eq.~\eqref{eq:dlog}. Quoted
uncertainties propagate the standard error of $q_4$.}
\label{tab:estimates}
\begin{ruledtabular}
\begin{tabular}{lccc}
Method & Input & $R$ & $s$\\
\hline
Ratios $r_1,r_2$ & $c_1,c_2$ & 8.557 & 1.426\\
Ratios $r_2,r_3$ & $c_1$--$c_3$ & 8.552 & 1.425\\
Ratios $r_3,r_4$ & $c_2$--$c_4$ & 8.55(7) & 1.42(3)\\
Dlog Pad\'e $[0/1]$ & $c_1,c_2$ & 8.557 & 1.426\\
Dlog Pad\'e $[1/1]$ & $c_1$--$c_3$ & 8.551 & 1.424\\
Dlog Pad\'e $[2/1]$ & $c_1$--$c_4$ & 8.54(10) & 1.42(5)\\
Dlog Pad\'e $[0/3]$ & $c_1$--$c_4$ & 8.54(10) & 1.42(5)\\
\end{tabular}
\end{ruledtabular}
\end{table}

\subsection{Dlog-Pad\'e and Pad\'e approximants}

The logarithmic derivative of a power-law singularity is a simple pole
with residue $-s$, which Pad\'e approximants represent exactly. From
Table~\ref{tab:coeffs},
\begin{align}
 \frac{d}{d\alpha}\ln\frac{m^*}{m}={}&0.166667+0.0194775\,\alpha
 +0.00227773\,\alpha^2\nonumber\\
 &+0.0002666(32)\,\alpha^3+\cdots .
 \label{eq:dlog}
\end{align}
The successive ratios of these coefficients are $0.116865$, $0.116942$,
and $0.1170(14)$: within the accuracy of the coefficients the series
\eqref{eq:dlog} is geometric, which is the statement that $\M$ is a pure
power of $(1-\alpha/R)$ through the order computed. The Dlog-Pad\'e
estimates of the pole and residue are listed in the lower part of
Table~\ref{tab:estimates}. (The $[1/2]$ approximant has a nearly
cancelling pole-zero pair and is omitted.) We summarize
Table~\ref{tab:estimates} as
\begin{equation}
 R=8.55(10),\qquad s=1.42(5),
 \label{eq:Rs}
\end{equation}
where the uncertainty is that of the fourth-order estimates and the
lower-order estimates agree far more closely. The corresponding
representation
\begin{equation}
 \frac{m^*}{m}\simeq\Bigl(1-\frac{\alpha}{R}\Bigr)^{-s}
 \label{eq:twopar}
\end{equation}
is more constraining than a two-parameter description of four numbers
might suggest. If $R$ and $s$ are fixed by the exactly known $c_1$ and $c_2$ alone
($R=8.5569$, $s=1.4261$), Eq.~\eqref{eq:twopar} predicts
\begin{equation}
 q_3=4.639\times10^{-5},\qquad q_4=2.13\times10^{-6},
 \label{eq:predq}
\end{equation}
to be compared with the computed $4.614\times10^{-5}$ and
$2.1(4)\times10^{-6}$. The third-order curvature is thus predicted to
$0.5\%$ and the fourth-order one to well within its uncertainty
[Fig.~\ref{fig:fourth}(b)]. For the next order the same form predicts
$q_5\simeq1.3\times10^{-7}$, i.e., $c_5\simeq5.2\times10^{-5}$.

Ordinary Pad\'e approximants to $\M$ itself can only mimic the branch
point by poles, and place their nearest pole somewhat inside the circle of
convergence: at $7.73$, $8.07$, and $8.10$ for the $[3/1]$, $[2/2]$, and
$[1/3]$ approximants, respectively. They are nevertheless accurate
representations of the function well inside the circle, as shown in
Sec.~\ref{sec:dmc}.

\subsection{Zeros of the truncated inverse mass: the fate of $\alpha=6$}

RS perturbation theory for the dispersion yields directly the series for
the \emph{inverse} mass. Its $N$th-order truncation,
\begin{equation}
 D_N(\alpha)=\sum_{n=0}^{N}d_n\alpha^n ,
 \label{eq:DN}
\end{equation}
is the object whose first-order zero is the familiar $\alpha=6$. [In the
language of approximation theory $1/D_N$ is the $[0/N]$ Pad\'e
approximant to $\M$; it should not be confused with Brillouin-Wigner
perturbation theory, which has self-consistent energy denominators.] With
the coefficients of Table~\ref{tab:coeffs},
\begin{align}
 D_4(\alpha)={}&1-\frac{\alpha}{6}+4.150147\times10^{-3}\alpha^2
 +9.2274\times10^{-5}\alpha^3\nonumber\\
 &+4.2(8)\times10^{-6}\alpha^4 .
\end{align}
The zeros of $D_N$ are listed in Table~\ref{tab:zeros}. For $N\ge2$ there
are several, and those not connected to the first-order zero move
erratically from order to order; they have no significance. The smallest
positive zero, which is the continuation of $\alpha=6$, forms the
monotonic sequence
\begin{equation}
 \alpha_N=6,\ 7.3424,\ 7.7565,\ 7.96(4)\qquad(N=1,\dots,4).
 \label{eq:alphaN}
\end{equation}
Equation~\eqref{eq:twopar} explains this sequence quantitatively. The
truncated Taylor polynomials of $(1-\alpha/R)^{s}$ with the values
\eqref{eq:Rs} have smallest positive zeros $6.004$, $7.340$, $7.755$, and
$7.958$, and these approach $R$ from below only as
$R-\alpha_N\simeq2.3/N$ [Fig.~\ref{fig:series}(b)]. The sequence
\eqref{eq:alphaN} is therefore nothing but a slowly convergent estimator
of the radius of convergence. The coupling $\alpha=6$ has no special
significance; the number that the perturbation series is trying to produce
is $R\simeq8.55$.

\begin{table}[t]
\caption{Zeros of the $N$th-order RS truncation $D_N(\alpha)$ of the
inverse mass, Eq.~\eqref{eq:DN}. The smallest positive zero $\alpha_N$ is
compared with that of the truncated series of $(1-\alpha/R)^s$ with
$R=8.55$, $s=1.424$.}
\label{tab:zeros}
\begin{ruledtabular}
\begin{tabular}{clcc}
$N$ & Other zeros & $\alpha_N$ & Model\\
\hline
1 & & 6 & 6.004\\
2 & $32.82$ & 7.3424 & 7.340\\
3 & $19.38$, $-72.11$ & 7.7565 & 7.755\\
4 & $15.5(5)$, $-23(3)\pm38(3)i$ & 7.96(4) & 7.958\\
\end{tabular}
\end{ruledtabular}
\end{table}

\subsection{Comparison with diagrammatic Monte Carlo}
\label{sec:dmc}

\begin{figure}[t]
 \centering
 \includegraphics[width=\columnwidth]{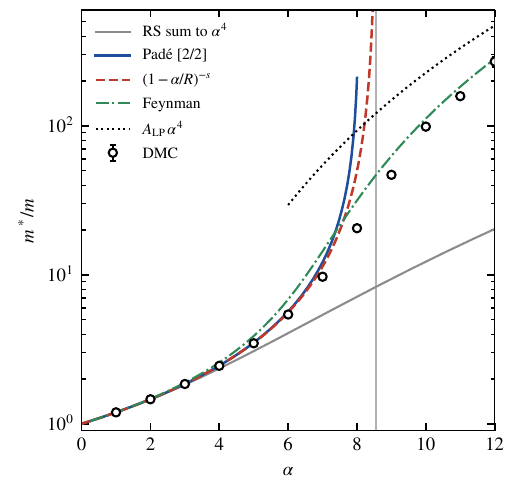}
 \caption{Effective mass versus coupling. Circles: DMC results of
 Ref.~\cite{Hahn2018}. Gray line: RS series summed through $\alpha^4$.
 Blue line: its $[2/2]$ Pad\'e approximant. Dashed: Eq.~\eqref{eq:twopar}
 with $R=8.55$, $s=1.424$. Dash-dotted: Feynman's variational mass.
 Dotted: the rigorous strong-coupling asymptote, Eq.~\eqref{eq:LP}. The
 vertical line marks $\alpha=R$.}
 \label{fig:mass}
\end{figure}

A series with radius of convergence $R\simeq8.5$ should describe the
polaron quantitatively well beyond the range $\alpha\lesssim1$ usually
assigned to weak-coupling theory, provided it is continued sensibly. This
is borne out by Table~\ref{tab:dmc} and Fig.~\ref{fig:mass}, where we
compare with the DMC results of Hahn \emph{et al.}~\cite{Hahn2018} and
with Feynman's variational theory \cite{Feynman1955,Schultz1959}.

\begin{table*}[t]
\caption{Ground-state energy and effective mass from the weak-coupling
series compared with diagrammatic Monte Carlo (DMC) and with Feynman's
variational theory (both from Ref.~\cite{Hahn2018}). ``RS'' is the
partial sum of the series through $\alpha^5$ (energy) or $\alpha^4$
(mass); ``Pad\'e'' is the $[2/2]$ approximant to $-E_0/\alpha$ or to
$\M$, with uncertainties propagated from $e_5$ or $q_4$ where they exceed
one unit in the last digit; the next column is Eq.~\eqref{eq:twopar} with
$R=8.55$, $s=1.424$. The $[2/2]$ approximant to the mass has a pole at
$\alpha=8.07$.}
\label{tab:dmc}
\begin{ruledtabular}
\begin{tabular}{c|llll|lllll}
 & \multicolumn{4}{c|}{$-E_0$} & \multicolumn{5}{c}{$\M$}\\
$\alpha$ & RS & Pad\'e & DMC & Feynman & RS & Pad\'e &
Eq.~\eqref{eq:twopar} & DMC & Feynman\\
\hline
1 & 1.0168 & 1.0168 & 1.01662(47) & 1.0130 & 1.1939 & 1.1939 & 1.1938 & 1.19396(2) & 1.1955\\
2 & 2.0711 & 2.0711 & 2.06957(84) & 2.0554 & 1.4596 & 1.4618 & 1.4615 & 1.46166(7) & 1.4719\\
3 & 3.1703 & 3.1705 & 3.1683(14) & 3.1333 & 1.8309 & 1.8511(3) & 1.8503 & 1.85047(13) & 1.8890\\
4 & 4.3238 & 4.3254(5) & 4.3249(21) & 4.2565 & 2.3510 & 2.458(2) & 2.4553 & 2.45196(57) & 2.5793\\
5 & 5.5439 & 5.551(2) & 5.5530(30) & 5.4401 & 3.0733 & 3.506(9) & 3.496 & 3.4719(18) & 3.8856\\
6 & 6.8458 & 6.869(7) & 6.8665(29) & 6.7109 & 4.0606 & 5.67(5) & 5.60 & 5.4195(63) & 6.8384\\
7 & 8.2484 & 8.31(2) & 8.3104(31) & 8.1127 & 5.3859 & 12.1(5) & 11.4 & 9.713(27) & 14.394\\
8 & 9.7744 & 9.95(6) & 9.9221(61) & 9.6954 & 7.1316 & & 49.8 & 20.55(14) & 31.569\\
\end{tabular}
\end{ruledtabular}
\end{table*}

For the mass, the $[2/2]$ Pad\'e approximant built from $c_1,\dots,c_4$
agrees with DMC to $0.03\%$ or better for $\alpha\le3$, to $0.2\%$ at
$\alpha=4$, to $1\%$ at $\alpha=5$, and to $5\%$ at $\alpha=6$; the
$[3/1]$ and $[1/3]$ approximants and Eq.~\eqref{eq:twopar} give the same
picture. Feynman's mass, by comparison, exceeds the DMC value by $2\%$,
$12\%$, and $26\%$ at $\alpha=3$, $5$, and $6$. The bare partial sum of
the series, which converges only slowly as $\alpha$ approaches $R$,
underestimates the mass by $25\%$ at $\alpha=6$, which illustrates why the
series has not usually been regarded as useful at such couplings. For
$\alpha\ge7$ all series-based estimates overshoot, as they must, since
each of them has a singularity on the real axis between $7.7$ and $8.6$
that the exact function does not have. For the energy, the $[2/2]$
approximant to $-E_0/\alpha$ built from $e_1,\dots,e_5$ agrees with DMC
within the combined uncertainties over the entire range of the table,
i.e., to better than $0.1\%$ for $\alpha\le7$ and to $0.3(6)\%$ at
$\alpha=8$, whereas Feynman's variational bound lies $1$--$2.4\%$ above
the DMC energy for $\alpha\ge3$.

The DMC data also show directly that $\alpha\simeq R$ is where the
polaron changes character. The logarithmic slope of the DMC mass,
$\Delta\ln m^*/\Delta\alpha$, between successive integer couplings from
$\alpha=6$ to $\alpha=11$ is $0.58$, $0.75$, $0.83$, $0.75$, and $0.47$:
it is largest between $\alpha=8$ and $9$ and decreases thereafter toward
the strong-coupling value $4/\alpha$. The radius of convergence of the
weak-coupling series coincides with the inflection of $\ln m^*$, that is,
with the center of the crossover from the weak-coupling to the
Landau-Pekar regime (Fig.~\ref{fig:mass}).

\section{What the series does not determine}
\label{sec:discussion}

\subsection{Distance of the singularity from the real axis}
\label{sec:distance}

The Proposition of Sec.~\ref{sec:signs} requires the singularity found in
Sec.~\ref{sec:series} to be off the real axis, and the DMC mass is indeed
finite and smooth at $\alpha=R$. The simplest representation compatible
with all of this information replaces Eq.~\eqref{eq:twopar} by
\begin{equation}
 \frac{m^*}{m}\simeq\Bigl[1-\frac{2\alpha}{R}\cos\theta
 +\frac{\alpha^2}{R^2}\Bigr]^{-s/2},
 \label{eq:pair}
\end{equation}
which has branch points at $\alpha_c=Re^{\pm i\theta}$. Its Taylor
coefficients are $R^{-n}C_n^{(s/2)}(\cos\theta)$, with $C_n^{(\lambda)}$
the Gegenbauer polynomials, and the coefficients of its logarithmic
derivative are $s\cos[(n+1)\theta]/R^{n+1}$, in accordance with
Eq.~\eqref{eq:darboux}. Two independent pieces of information constrain
$\theta$, and they do not agree.

The first is the regularity of the computed coefficients. If $R$ and $s$
are fixed at given $\theta$ by $c_1$ and $c_2$, Eq.~\eqref{eq:pair}
predicts $q_3=4.64$, $4.73$, and $5.03\times10^{-5}$ for $\theta=0$,
$0.05$, and $0.10$, against the computed $4.614\times10^{-5}$. The
coefficients thus favor $\theta\lesssim0.03$. The second is the size of
the mass in the crossover region. On the real axis Eq.~\eqref{eq:pair}
reaches $\M\simeq\theta^{-s}$ at $\alpha\simeq R$, with a maximal
logarithmic slope $s/(2R\theta)$. The DMC mass interpolated to
$\alpha=8.55$ is about $32$ and the maximal DMC slope is $0.83$; these
give $\theta\simeq0.09$ and $\theta\simeq0.10$, respectively.

We conclude that the minimal one-pair, pure-power model \eqref{eq:pair}
is too crude: the true analytic structure near $|\alpha|\simeq8.5$ must
involve more than a single conjugate pair of algebraic branch points, for
instance confluent corrections or further nearby singularities, whose
contributions largely cancel in the first few coefficients. This is
precisely the information that four coefficients do not contain. What can
be said robustly is that the modulus of the nearest singularity is fixed,
$|\alpha_c|\simeq8.5$, while its angle is small and undetermined,
$0<\theta\lesssim0.1$.

This has a sobering consequence for the prospects of going further by
brute force. The first negative coefficient of Eq.~\eqref{eq:pair} occurs
at order $n^*\simeq(1+s/2)\pi/(2\theta)\simeq2.7/\theta$; explicitly,
$n^*=27$, $54$, and $91$ for $\theta=0.10$, $0.05$, and $0.03$. Until
orders of this magnitude are reached, the coefficients of the polaron
series will presumably continue to look like those of a function with a
divergence on the real axis, i.e., like those of a model with a
self-trapping transition at $\alpha\simeq8.5$, even though no such
transition exists. The number of proper histories at order $n$ grows as
$1,2,10,74,706,8162,\dots$, so that the direct evaluation of the
coefficients at $n\sim30$ is out of the question. A few additional
orders would sharpen $R$ and $s$ and test the predictions following
Eq.~\eqref{eq:predq}, but would not reveal how the divergence is rounded.

\subsection{Strong coupling, Borel methods, and two-point interpolation}
\label{sec:borel}

The strong-coupling regime, $\alpha\gg R$, lies outside the circle of
convergence, and the Landau-Pekar law \eqref{eq:LP} is a statement about
the analytic continuation of the series along the positive real axis past
the pair of singularities at $|\alpha|\simeq8.5$. None of the
continuations considered above can reach it: the truncations $1/D_N$ decay
as $\alpha^{-N}$, the Pad\'e approximants of Table~\ref{tab:dmc} have
poles on the real axis, and Eq.~\eqref{eq:twopar} ceases to be real
beyond $R$.

We have also examined Borel-Leroy-Pad\'e continuations, in which Pad\'e
approximants are applied to
$\sum_nc_nz^n/\Gamma(n+b+1)$ and the result is Laplace transformed. For a
convergent series this procedure has no justification, since the function
being approximated by rational functions is entire, and it performs
accordingly. The $[3/1]$ approximant has a pole on the positive axis for
every Leroy parameter tried ($b=0,1,2,4$), the $[1/3]$ approximant for
$b=0,1,2$, and the pole-free $[2/2]$ approximant gives
$\M=3.34$--$3.69$ at $\alpha=6$, depending on $b$. These values lie below
even the partial sum of the series, $4.06$, all of whose terms are
positive, and far below the DMC value $5.42$, whereas ordinary Pad\'e
approximants give $5.65$--$5.89$. Borel methods should not be applied to
this problem.

One may, finally, impose the strong-coupling law as an additional input
and interpolate between the two limits, as has been done in various ways
\cite{SelyuginSmondyrev1989,Kleinert1995}. Such two-point constructions
can be useful, but they are not determined by the endpoint data: adding to
$\M$ any term that is $\OO(\alpha^5)$ at small $\alpha$ and
$o(\alpha^4)$ at large $\alpha$, such as $\gamma\alpha^5/(1+\alpha)^2$,
changes neither the known weak-coupling coefficients nor the
Landau-Pekar amplitude but alters the crossover at will. The results of
this paper suggest a more constrained strategy. An interpolant should
reproduce not only the endpoint data but also the location
$|\alpha_c|\simeq8.5$ of the nearest singularities, should be analytic on
the positive real axis, and should be tested against the DMC data of
Table~\ref{tab:dmc} rather than assumed. We leave this to future work.

We have carried out the same fourth-order calculation and series analysis
for the three-dimensional Holstein polaron, where the coupling is local
rather than long ranged (Appendix~\ref{app:holstein}). There are
substantial quantitative differences: the Holstein coefficients are much
less regular than the Fr\"ohlich ones, the singularity they point to is far
weaker, and its relation to the crossover depends on the adiabaticity
ratio. The qualitative conclusions reached above with Pad\'e and
Borel-Pad\'e continuations are nevertheless the same for the two models, at
least with the fourth-order results available: the weak-coupling series
converges, ordinary Pad\'e approximants reproduce the numerically exact
mass to within a few percent up to roughly $70\%$ of the crossover coupling
and fail beyond it, Borel-Pad\'e continuation does worse than the bare
partial sums, and the weak-coupling coefficients do not determine how the
polaron passes through the crossover.

\section{Conclusions}
\label{sec:conclusions}

We have extended the Rayleigh-Schr\"odinger series for the effective mass
of the three-dimensional Fr\"ohlich polaron from second to fourth order,
Eq.~\eqref{eq:massseries}, and have confirmed the known fourth-order
energy coefficient by an independent method. The precision of the
fourth-order curvature coefficient, about $20\%$, is limited by sharp
features of the nine-dimensional integrand (Appendix~\ref{app:numerics})
and can be improved.

The series has a nonzero radius of convergence, and its first four
coefficients determine that radius: $R=8.55(10)$, with the mass behaving
as $(1-\alpha/R)^{-1.42(5)}$ within the circle. The well-known zero of the
leading-order inverse mass at $\alpha=6$ is the first term of a sequence,
$6\to7.34\to7.76\to7.96$, that converges slowly to $R$. Continued by
ordinary Pad\'e approximants, the weak-coupling series reproduces
numerically exact Monte Carlo masses to $1\%$ up to $\alpha=5$ and
energies to $0.1\%$ up to $\alpha=7$, a far wider range than is usually
credited to perturbation theory.

Exact results forbid a singularity on the real axis and require the
coefficients of both the energy and the mass series to change sign
infinitely often, which none of the known coefficients do. The
singularity at $|\alpha|\simeq8.5$ is therefore complex, close to the real
axis, and is the analytic signature of the smooth but rapid crossover to
the Landau-Pekar regime. The weak-coupling expansion determines the
location of this crossover with an accuracy of about one percent. It does
not determine its width, and the coefficients that would do so lie at
orders far beyond direct calculation. The connection between weak and
strong coupling in the polaron problem is thus limited not by the
divergence of perturbation theory, as in QED, but by a pair of complex
singularities that a convergent perturbation series can locate but not
cross. A parallel fourth-order analysis of the three-dimensional Holstein
polaron (Appendix~\ref{app:holstein}) leads to the same qualitative
conclusions, with a markedly less regular series.

\begin{acknowledgments}
This work is supported by the Laboratory for Physical Sciences. The large
language models GPT-5.6 Terra Extra High and Claude Opus 5.5 were used to assist
with the numerical calculations, the series analysis, the figures, and the
preparation of the manuscript; the author takes full responsibility for
the content.
\end{acknowledgments}

\section*{Data availability}
The data that support the findings of this article are contained in the
article, in Tables~\ref{tab:coeffs}, \ref{tab:resolution},
\ref{tab:raw}, and \ref{tab:hcoef}--\ref{tab:hzeros}.

\appendix
\section{Numerical evaluation at fourth order}
\label{app:numerics}

\emph{Integration variables.} Rotational invariance allows $\bm q_1$ to be
placed along the $z$ axis and $\bm q_2$ in the $xz$ plane, which leaves
nine variables: four radii, three polar cosines, and two azimuths. Each
radius is mapped to the unit interval by
\begin{equation}
 q=\sqrt2\tan\frac{\pi u}{2},\qquad 0<u<1 ,
\end{equation}
which absorbs the vertex factor and the radial measure, since
$\int_{\bm q}|v_{\bm q}|^2(\cdots)=\int_0^1du\,(1+q^2/2)
\langle\cdots\rangle_{\hat{\bm q}}$. The integrand in the unit hypercube
is bounded.

\emph{Sampling and errors.} The nine-dimensional integral is evaluated
with base-2 Sobol nets with independent random digital scrambles
\cite{Sobol1967,Owen1995}. The quoted uncertainties are standard errors
of the mean over independent scrambles and are not inferred from the
convergence of a single net. Table~\ref{tab:resolution} shows the
dependence on the number of points and Table~\ref{tab:raw} lists the nine
runs entering Eq.~\eqref{eq:e4q4}, which are also shown in
Fig.~\ref{fig:fourth}.

\emph{Limitation.} The scatter between scrambles decreases only slowly
between $2^{23}$ and $2^{24}$ points, more slowly than would be expected
for a smooth integrand. The reason is that the integrand, although
bounded, has narrow ridges at large momenta where two or more phonon
momenta nearly cancel, $\bm q_i+\bm q_j\simeq0$, so that an intermediate
denominator in Eq.~\eqref{eq:history} remains of order unity while the
individual momenta are large; the product-form radial map does not
resolve these regions. Symmetrization of the integrand over permutations
of the momenta, which improves the pointwise cancellation among the
monomials of Eq.~\eqref{eq:e4}, and an importance map adapted to the
ridges should both reduce the variance substantially.

\begin{table}[t]
\caption{Fourth-order results versus the number of Sobol points per
scramble. The last line pools the nine runs with $2^{23}$ and $2^{24}$
points and is the result quoted in Eq.~\eqref{eq:e4q4}.}
\label{tab:resolution}
\begin{ruledtabular}
\begin{tabular}{cccc}
Points & Scrambles & $10^5e_4$ & $10^6q_4$\\
\hline
$2^{20}$ & 9 & $-5.52(46)$ & $-1.6(6.8)$\\
$2^{22}$ & 8 & $-5.60(17)$ & $1.2(1.2)$\\
$2^{23}$ & 6 & $-5.22(13)$ & $2.09(53)$\\
$2^{24}$ & 3 & $-5.47(15)$ & $2.14(65)$\\
\hline
$2^{23}$, $2^{24}$ & 9 & $-5.31(10)$ & $2.11(39)$\\
\end{tabular}
\end{ruledtabular}
\end{table}

\begin{table}[t]
\caption{The nine independent scrambled-Sobol runs entering
Eq.~\eqref{eq:e4q4}; $2^m$ is the number of points.}
\label{tab:raw}
\begin{ruledtabular}
\begin{tabular}{cccc}
$m$ & Seed & $10^5e_4$ & $10^6q_4$\\
\hline
23 & 7103 & $-4.6894$ & $4.1656$\\
23 & 7207 & $-5.4454$ & $2.0580$\\
23 & 7301 & $-5.3763$ & $0.4896$\\
23 & 7409 & $-5.5672$ & $2.0498$\\
23 & 7503 & $-5.1683$ & $2.7048$\\
23 & 7607 & $-5.0897$ & $1.0785$\\
24 & 7703 & $-5.6823$ & $1.8408$\\
24 & 7807 & $-5.1950$ & $3.3733$\\
24 & 7901 & $-5.5469$ & $1.1909$\\
\end{tabular}
\end{ruledtabular}
\end{table}

\section{Comparison with the three-dimensional Holstein polaron}
\label{app:holstein}

To see how much of the picture developed in the main text is specific to the
long-range Fr\"ohlich coupling, we have carried out the same fourth-order
calculation and the same series analysis for the opposite limit of a strictly
local coupling, the Holstein model~\cite{Holstein1959} of one electron on a
simple cubic lattice,
\begin{equation}
H=-t\sum_{\langle ij\rangle}c_i^\dagger c_j^{}
  +\omega_0\sum_i b_i^\dagger b_i^{}
  +g\sum_i c_i^\dagger c_i^{}\,(b_i^{}+b_i^\dagger).
\label{eq:holstein}
\end{equation}
The model has two dimensionless parameters, the adiabaticity ratio
$\omega_0/t$ and the coupling
\begin{equation}
\lambda=\frac{g^2}{6t\,\omega_0},
\end{equation}
which is the ratio of the atomic-limit polaron energy $g^2/\omega_0$ to half
the bare bandwidth. With the lattice constant set to one, the bare band is
$\epsilon_{\bm k}=-2t\sum_{i}\cos k_i$ and the band mass is $m=1/2t$. In
analogy with Eq.~\eqref{eq:series} we write
\begin{eqnarray}
E_0&=&-6t+t\sum_{n\ge1}e_n\lambda^n,\qquad
\frac{m^*}{m}=\sum_{n\ge0}c_n\lambda^n,\nonumber\\
\frac{m}{m^*}&=&\sum_{n\ge0}d_n\lambda^n ,
\label{eq:hseries}
\end{eqnarray}
with $c_0=d_0=1$; order $n$ is order $g^{2n}$, as order $\alpha^n$ is in the
Fr\"ohlich model.

\subsection{Convergence and evaluation of the series}

The argument of Sec.~\ref{sec:convergence} applies without change, and in fact
more simply. At fixed total momentum the perturbation is the phonon field
operator at a single site, which is infinitesimally small with respect to the
phonon number operator, and the unperturbed state at zero total momentum is
nondegenerate and separated by $\omega_0$ from the one-phonon threshold. The
Hamiltonian is therefore a holomorphic family in $g$~\cite{Kato1976}, the
ground-state energy is even in $g$, and the series \eqref{eq:hseries} have
nonzero radii of convergence for every $\omega_0>0$. As in the Fr\"ohlich
case, the ground state is analytic for all real coupling, so that there is no
self-trapping transition at any
$\omega_0>0$~\cite{Lowen1988,GerlachLowen1991}.

The RS brackets are sums over the same proper histories as in
Sec.~\ref{sec:histories} (Fig.~\ref{fig:diagrams}), namely 1, 2, 10, and 74
at orders one to four. Because the Holstein phonon is local, the two vertices
joined by a contraction sit on the same lattice site, and a history with $k$
phonons becomes a sum over $k-1$ lattice vectors of a product of $2k-1$
lattice Green functions
\begin{equation}
G_{\bm r}(\Omega)=\frac1N\sum_{\bm p}
\frac{e^{i\bm p\cdot\bm r}}{\Omega+\epsilon_{\bm p}-\epsilon_{\bm 0}} .
\end{equation}
It is convenient to organize the calculation in Brillouin-Wigner form. With
$\eta(\bm k)=E(\bm k)-\epsilon_{\bm 0}$ one has, exactly,
$\eta=\epsilon_{\bm k}-\epsilon_{\bm 0}+\Sigma(\bm k,\eta)$ with
\begin{align}
\Sigma(\bm k,\eta)={}&-\sum_{n\ge1}g^{2n}\!\sum_{h\in\mathcal H_n}\sum_{\{\bm s\}}
e^{i\bm k\cdot(\bm s_1-\bm s_{2n})}\nonumber\\
&\times\prod_{j=1}^{2n-1}G_{\bm s_{j+1}-\bm s_j}\bigl(|S_j|\,\omega_0-\eta\bigr),
\label{eq:hsigma}
\end{align}
where $\bm s_j$ is the site of the $j$th vertex and the sum runs over the
positions of the $n$ phonons relative to the first one; solving for $\eta$
as a power series in $g^2$ reproduces the RS coefficients of
Eqs.~\eqref{eq:e123} and \eqref{eq:e4}.
The effective mass follows from
\begin{equation}
\frac{m^*}{m}=\frac{Z^{-1}}{K},\quad
Z^{-1}=1-\frac{\partial\Sigma}{\partial\eta},\quad
K=1+\frac{\partial\Sigma}{\partial(tk^2)},
\label{eq:hmass}
\end{equation}
evaluated at $\bm k=0$ and $\eta=\eta(0)$. The momentum derivative acts
only on the phase factor in Eq.~\eqref{eq:hsigma} and amounts to a weight
$-|\bm s_1-\bm s_{2n}|^2/6t$. All lattice sums are evaluated on
periodic lattices of $L^3$ sites by fast Fourier transformation, with $L$ up
to 48. Finite-size corrections decay as $e^{-\kappa L}$ with
$\cosh\kappa=1+\omega_0/2t$, and between the two largest lattices used the
coefficients in Table~\ref{tab:hcoef} change by less than one part in $10^{5}$
at $\omega_0=0.2t$ and by less than one part in $10^{7}$ for
$\omega_0\ge0.5t$. There is no stochastic integration and hence no
statistical error: all four orders are known to the digits shown, in contrast
to the Fr\"ohlich coefficient $q_4$ of Eq.~\eqref{eq:e4q4}.

We have checked the calculation in three ways: against the exact ground-state
energies of periodic clusters of up to eight sites at every total momentum
below the one-phonon threshold, where the difference from the fourth-order
series scales as $g^{10}$; against the atomic limit including its first
correction in $t/\omega_0$, $c_n\to(6t/\omega_0)^n[1-12t/(n\omega_0)]/n!$; and
against the finite-momentum dispersion on lattices with $L=12$--$20$, which
confirms Eq.~\eqref{eq:hmass} order by order.

At first order $\Sigma$ is independent of $\bm k$ and
\begin{equation}
c_1\lambda=\frac{g^2}{N}\sum_{\bm p}
\frac{1}{(\omega_0+\epsilon_{\bm p}-\epsilon_{\bm 0})^{2}},
\end{equation}
which tends to $(3/4\pi)\lambda\,(\omega_0/t)^{1/2}$ for $\omega_0\ll t$ and
to $g^2/\omega_0^2$ for $\omega_0\gg t$. The self-energy remains nearly local
at higher orders. Its momentum dependence first appears at order $\lambda^2$,
through the crossed contraction, and the second-order coefficient of $K$ is
between $0.4\%$ and $2\%$ of that of $Z^{-1}$ for the values of $\omega_0/t$
in Table~\ref{tab:hcoef}. The Holstein mass enhancement is thus essentially
the inverse quasiparticle weight, consistent with the known accuracy of the
dynamical mean-field solution of the
model~\cite{Ciuchi1997,Mitric2022}, in which the self-energy is local. The
Fr\"ohlich polaron is quite different in this respect: there $Z^{-1}=1+\alpha/2+0.148\,\alpha^2$ and
$K=1+\alpha/3+0.069\,\alpha^2$ through second order, so that two thirds of the
leading enhancement of $Z^{-1}$ is cancelled by the momentum dependence of the
self-energy.

\begin{table*}[t]
\caption{Coefficients of the RS series \eqref{eq:hseries} for the ground-state
energy and the effective mass of the three-dimensional Holstein polaron on the
simple cubic lattice. All digits shown are significant.}
\label{tab:hcoef}
\begin{ruledtabular}
\begin{tabular}{ccccccccc}
$\omega_0/t$ & $e_1$ & $e_2$ & $e_3$ & $e_4$ & $c_1$ & $c_2$ & $c_3$ & $c_4$ \\
\hline
0.2 & $-0.25873$ & $-0.01304$ & $-0.002327$ & $-0.000645$ & $0.11331$ & $0.02578$ & $0.007859$ & $0.002893$ \\
0.5 & $-0.58162$ & $-0.04939$ & $-0.01394$ & $-0.005937$ & $0.17456$ & $0.06156$ & $0.02928$ & $0.01669$ \\
1 & $-1.02314$ & $-0.11195$ & $-0.03788$ & $-0.01851$ & $0.23013$ & $0.10292$ & $0.05999$ & $0.04043$ \\
2 & $-1.69118$ & $-0.19806$ & $-0.06424$ & $-0.02800$ & $0.27932$ & $0.13701$ & $0.08119$ & $0.05227$ \\
4 & $-2.57153$ & $-0.25188$ & $-0.05740$ & $-0.01522$ & $0.29855$ & $0.13167$ & $0.06257$ & $0.02961$ \\
\end{tabular}
\end{ruledtabular}
\end{table*}

\subsection{Comparison with exact diagonalization}

\begin{figure}[t]
 \centering
\includegraphics[width=\columnwidth]{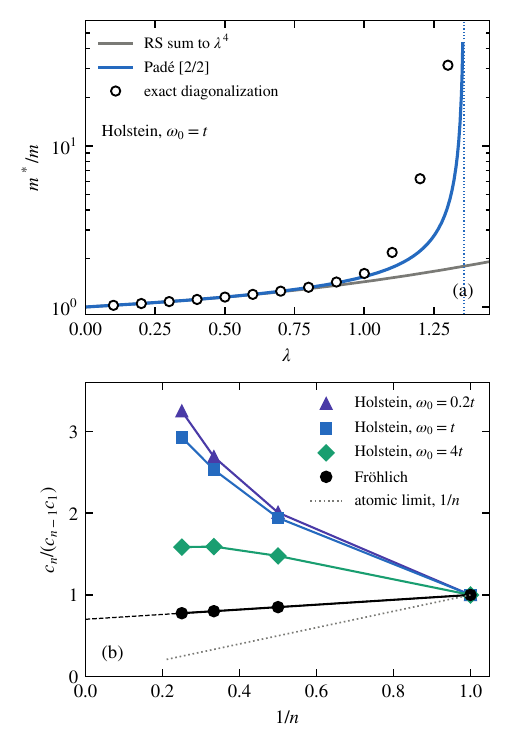}
\caption{(a) Effective mass of the three-dimensional Holstein polaron at
$\omega_0=t$. Circles: variational exact diagonalization. Gray line: RS series
summed through $\lambda^4$. Blue line: its $[2/2]$ Pad\'e approximant, whose
pole at $\lambda=1.36$ is marked by the dotted line. (b) Ratios of successive
mass coefficients, normalized by $c_1$, versus $1/n$ for the Fr\"ohlich
polaron (circles; the dashed line is Eq.~\eqref{eq:ratio} with $R=8.55$,
$s=1.424$) and for the Holstein polaron at three values of $\omega_0/t$. The
dotted line is the atomic limit $m^*/m=\exp(g^2/\omega_0^2)$, for which the
normalized ratio is $1/n$.}
\label{fig:holstein}
\end{figure}

Numerically exact results for the three-dimensional Holstein polaron are
available from variational exact diagonalization~\cite{Ku2002} and from
continuous-time quantum Monte Carlo~\cite{Kornilovitch1998}. To have a
benchmark at exactly the parameters of the series we have computed $E_0$ and
$m^*$ at $\omega_0=t$ by variational exact diagonalization in the
translationally invariant basis of Refs.~\cite{Bonca1999,Ku2002}, with
between $1.1\times10^6$ and $4.9\times10^6$ states generated by four
different truncation rules, the mass being obtained from $E(\bm k)$ at
small $k$. For $\lambda\le1$, where more than one basis was used they agree
to better than $0.1\%$ in the mass; for $1.1\le\lambda\le1.3$ the three bases
that allow the largest phonon numbers agree to $0.2\%$ in the mass and to
$2\times10^{-4}t$ in the energy. Table~\ref{tab:hexact} and
Fig.~\ref{fig:holstein}(a) compare these results with the series.

\begin{table}[t]
\caption{Holstein polaron at $\omega_0=t$: fourth-order series compared with
variational exact diagonalization (ED). ``RS'' is the partial sum through
$\lambda^4$; ``Pad\'e'' is the $[2/1]$ approximant to $-(E_0+6t)/\lambda t$
and the $[2/2]$ approximant to $m^*/m$, the latter having a pole at
$\lambda=1.36$. $N_{\rm ph}$ is the mean number of phonons in the ground
state.}
\label{tab:hexact}
\begin{ruledtabular}
\begin{tabular}{cccccccc}
 & \multicolumn{3}{c}{$-(E_0+6t)/t$} & \multicolumn{3}{c}{$m^*/m$} & \\
$\lambda$ & RS & Pad\'e & ED & RS & Pad\'e & ED & $N_{\rm ph}$ \\
\hline
0.2 & 0.2094 & 0.2094 & 0.2094 & 1.0507 & 1.0507 & 1.0507 & 0.05 \\
0.4 & 0.4301 & 0.4302 & 0.4302 & 1.1134 & 1.1138 & 1.1138 & 0.12 \\
0.6 & 0.6648 & 0.6658 & 0.6662 & 1.1933 & 1.1973 & 1.1977 & 0.21 \\
0.8 & 0.9171 & 0.9220 & 0.9252 & 1.2972 & 1.3201 & 1.3250 & 0.36 \\
0.9 & 1.0513 & 1.0608 & 1.0690 & 1.3607 & 1.4109 & 1.4284 & 0.49 \\
1.0 & 1.1915 & 1.2092 & 1.2311 & 1.4335 & 1.5421 & 1.6137 & 0.77 \\
1.1 & 1.3384 & 1.3700 & 1.4363 & 1.5167 & 1.7594 & 2.1789 & 1.68 \\
1.2 & 1.4928 & 1.5473 & 1.7687 & 1.6119 & 2.2241 & 6.24 & 4.47 \\
1.3 & 1.6554 & 1.7475 & 2.2587 & 1.7204 & 4.2207 & 31.6 & 6.63 \\
\end{tabular}
\end{ruledtabular}
\end{table}

The $[2/2]$ approximant to the mass agrees with the exact result to about
$0.1\%$ or better for $\lambda\le0.7$, to $0.4\%$ at $\lambda=0.8$, to $1.2\%$ at
$\lambda=0.9$, and to $4\%$ at $\lambda=1$; it is too small by $19\%$ at
$\lambda=1.1$ and by factors of $2.8$ and $7.5$ at $\lambda=1.2$ and $1.3$.
The exact mass has its steepest logarithmic growth between $\lambda=1.2$ and
$1.3$, where it increases fivefold, so that the crossover is centered at
$\lambda_\times\simeq1.25$ and is far more abrupt than in the Fr\"ohlich
model. Measured in units of the crossover coupling, the reach of the Pad\'e
approximant is comparable in the two models: one percent accuracy up to
$\lambda\simeq0.7\,\lambda_\times$ here, and up to $\alpha\simeq0.6R$ in
Table~\ref{tab:dmc}.

\subsection{Series analysis and convergence properties}

Tables~\ref{tab:hRs} and \ref{tab:hzeros} apply the methods of
Sec.~\ref{sec:series} to the Holstein coefficients and repeat the Fr\"ohlich results for
comparison; the coefficient ratios are shown in Fig.~\ref{fig:holstein}(b).
Several features are common to the two models. Both series converge. All
computed coefficients of $m^*/m$ are positive and all computed energy
coefficients are negative. In both models the coefficients point to a
singularity at a coupling of the order of the crossover coupling, and Pad\'e
approximants built from the weak-coupling coefficients alone are quantitatively
accurate over most of the weak-coupling side of the crossover and fail on
approaching it. The convergence properties of the two series are nevertheless
quite different, in the following respects.

\begin{table*}[t]
\caption{Ratio-method and Dlog-Pad\'e estimates of the location $R$ and
exponent $s$ of the apparent singularity of $m^*/m$, for the Fr\"ohlich
polaron (in the variable $\alpha$) and the Holstein polaron (in the variable
$\lambda$). The three ratio-method entries use the pairs $(r_1,r_2)$,
$(r_2,r_3)$, and $(r_3,r_4)$ in Eq.~\eqref{eq:ratio}; the two Dlog-Pad\'e
entries are the $[1/1]$ and $[2/1]$ approximants. The uncertainty of the
fourth-order Fr\"ohlich entries is given in Table~\ref{tab:estimates}; the Holstein
entries have no numerical uncertainty.}
\label{tab:hRs}
\begin{ruledtabular}
\begin{tabular}{lccccccc}
 & \multicolumn{3}{c}{$r_n/r_1$} & \multicolumn{2}{c}{Ratio method} & \multicolumn{2}{c}{Dlog Pad\'e} \\
 & $n=2$ & $n=3$ & $n=4$ & $R$ & $s$ & $R$ & $s$ \\
\hline
Fr\"ohlich ($\alpha$) & 0.851 & 0.801 & 0.776 & 8.557, 8.552, 8.55 & 1.426, 1.425, 1.42 & 8.551, 8.54 & 1.424, 1.42 \\
Holstein, $\omega_0/t=0.2$ & 2.008 & 2.691 & 3.249 & 2.93, 2.18, 1.79 & $0.33$, $-0.01$, $-0.36$ & 2.38, 2.08 & 0.22, 0.15 \\
Holstein, $\omega_0/t=0.5$ & 2.020 & 2.725 & 3.266 & 1.88, 1.39, 1.17 & $0.33$, $-0.02$, $-0.33$ & 1.52, 1.34 & 0.21, 0.15 \\
Holstein, $\omega_0/t=1$ & 1.943 & 2.533 & 2.929 & 1.51, 1.17, 1.06 & $0.35$, $0.05$, $-0.15$ & 1.26, 1.16 & 0.24, 0.19 \\
Holstein, $\omega_0/t=2$ & 1.756 & 2.121 & 2.305 & 1.43, 1.26, 1.25 & $0.40$, $0.23$, $0.23$ & 1.30, 1.28 & 0.33, 0.32 \\
Holstein, $\omega_0/t=4$ & 1.477 & 1.592 & 1.585 & 1.71, 1.84, 2.14 & $0.51$, $0.62$, $1.05$ & 1.81, 2.01 & 0.57, 0.78 \\
\end{tabular}
\end{ruledtabular}
\end{table*}

\begin{table}[t]
\caption{Smallest positive zero of the $N$th-order RS truncation of the
inverse mass, Eq.~\eqref{eq:DN}, the pole of the $[2/2]$ Pad\'e approximant
to $m^*/m$, and the coupling at which $\ln(m^*/m)$ grows most steeply (the
center of the crossover). Fr\"ohlich entries are values of $\alpha$, Holstein
entries values of $\lambda$.}
\label{tab:hzeros}
\begin{ruledtabular}
\begin{tabular}{lcccccc}
 & $N=1$ & 2 & 3 & 4 & Pad\'e & Crossover \\
\hline
Fr\"ohlich ($\alpha$) & 6.00 & 7.34 & 7.76 & 7.96 & 8.07 & 8--9 \\
Holstein, $\omega_0/t=0.2$ & 8.83 & 5.44 & 4.29 & 3.68 & 2.39 & $\simeq 0.9$--$1.0$\footnotemark[1] \\
Holstein, $\omega_0/t=0.5$ & 5.73 & 3.52 & 2.77 & 2.37 & 1.56 & 1.03\footnotemark[2] \\
Holstein, $\omega_0/t=1$ & 4.35 & 2.73 & 2.18 & 1.91 & 1.36 & 1.25 \\
Holstein, $\omega_0/t=2$ & 3.58 & 2.38 & 1.99 & 1.80 & 1.49 & 1.75\footnotemark[2] \\
Holstein, $\omega_0/t=4$ & 3.35 & 2.48 & 2.24 & 2.16 & 2.11 & $\simeq 2.8$\footnotemark[2] \\
\end{tabular}
\footnotetext[1]{The adiabatic energy functional has a first-order
self-trapping transition at $\lambda_c=0.90$~\cite{Kabanov1993}.}
\footnotetext[2]{From the local-self-energy (dynamical mean-field)
approximation of Ref.~\cite{Ciuchi1997}, evaluated with the simple cubic
density of states; at $\omega_0=t$ it agrees with the exact mass to $5\%$ or
better for $\lambda\le1.3$.}
\end{ruledtabular}
\end{table}

\emph{Regularity.} The Fr\"ohlich coefficients follow the two-parameter form
\eqref{eq:twopar} so closely that $R$ and $s$ are stable to one part in $10^3$
from one order to the next. Nothing comparable holds for the Holstein model.
At $\omega_0=t$ the successive ratio estimates of $R$ are $1.51$, $1.17$, and
$1.06$, and those of $s$ are $0.35$, $0.05$, and $-0.15$; the Dlog-Pad\'e
estimates drift in the same direction. Four exactly known coefficients locate
the apparent singularity only to about $15\%$; the fourth-order estimates,
$R\simeq1.06$--$1.20$, lie just below the crossover coupling
$\lambda_\times\simeq1.25$. Only for $\omega_0\simeq2t$ are the estimates
stable ($R\simeq1.25$--$1.30$, $s\simeq0.2$--$0.3$), and there $R$ lies well below the center
of the crossover.

\emph{Nature of the singularity.} The Fr\"ohlich ratios decrease with $n$
toward $1/R$, as for a divergence with $s>1$, whereas the Holstein ratios
increase with $n$ for $\omega_0\lesssim2t$ [Fig.~\ref{fig:holstein}(b)], which
corresponds to $s<1$; the estimates give $s\simeq0.2\pm0.2$ at $\omega_0=t$.
The apparent Holstein singularity is therefore not a divergence of the mass
of the kind suggested by the Fr\"ohlich coefficients, but a much weaker
singularity, across which the mass nevertheless grows by two orders of
magnitude. For $\omega_0\gg t$ the ratios turn over and approach the $1/n$
behavior of the atomic limit, for which the series is that of the entire
function $\exp(g^2/\omega_0^2)$.

\emph{The first-order criterion.} In the Fr\"ohlich model the zeros of the
truncated inverse mass increase toward $R$ from below, starting from
$\alpha=6$, which is $30\%$ too small. In the Holstein model they decrease
from above, starting from $\lambda_1=1/c_1$, which at $\omega_0=t$ is $4.35$,
three and a half times the crossover coupling, and they are still at $1.9$ in
fourth order. The vanishing of the leading-order inverse mass is thus not
even a rough guide to the breakdown of weak coupling in the Holstein model.

\emph{Failure of the continuation.} Beyond the range in which they are
accurate, the Pad\'e approximants fail in opposite directions. The Fr\"ohlich
approximants overshoot, because they place a pole on the real axis just below
a crossover that is in fact smooth (Table~\ref{tab:dmc}). The Holstein
approximants undershoot, by a factor of $7.5$ at $\lambda=1.3$, because their
pole lies above a crossover that is far more abrupt than the low-order
coefficients indicate.

\emph{Dependence on $\omega_0/t$.} The Fr\"ohlich problem has a single
coupling constant, so that the scale on which the series varies and the scale
of the crossover are necessarily the same. In the Holstein model they
separate in the adiabatic regime. For $\omega_0\le t$ the normalized
coefficients $c_n/c_1^{\,n}$ depend only weakly on $\omega_0/t$ ($1.97$,
$5.0$, $15$ at $\omega_0=0.1t$; $2.02$, $5.5$, $18$ at $\omega_0=0.5t$), so
that the series is a function of the single variable
$c_1\lambda\simeq(3/4\pi)\lambda(\omega_0/t)^{1/2}$, with an apparent
singularity at $c_1\lambda\simeq0.2$--$0.25$. The crossover, on the other
hand, stays at $\lambda$ of order one and sharpens as $\omega_0/t$ decreases,
approaching the first-order self-trapping transition of the adiabatic
theory, which for the simple cubic lattice occurs at $g^2/\omega_0=5.41\,t$,
i.e., $\lambda_c=0.90$ (Ref.~\cite{Kabanov1993}; we reproduce this value). The apparent radius of convergence in $\lambda$ therefore grows
roughly as $(t/\omega_0)^{1/2}$ while the crossover does not move
(Tables~\ref{tab:hRs} and \ref{tab:hzeros}): at $\omega_0=0.2t$ the fourth-order estimates are
$R\simeq1.8$--$2.1$, twice the crossover coupling, and still decreasing with
the order. The low-order coefficients describe the weakly dressed state and
carry no visible trace of the avoided level crossing by which the ground
state leaves it. Since the mass changes by orders of magnitude within a
narrow interval of real $\lambda$, there must be singularities close to the
real axis near $\lambda\simeq1$; their weight in the low-order coefficients
is evidently very small, so that they can control the series only at high
order. This is the
same obstruction as in Sec.~\ref{sec:distance}, in a more severe form: for the
Fr\"ohlich polaron four coefficients fix the location of the crossover but not
its width, whereas for the adiabatic Holstein polaron they do not fix its
location either.

\emph{Sign of the coefficients.} The Proposition of Sec.~\ref{sec:signs}
carries over only in part. Since the ground state is analytic for all real
$g$, a finite radius of convergence again implies, by the Vivanti-Pringsheim
theorem~\cite{Titchmarsh1939}, that the nearest singularities are complex and
that the coefficients cannot ultimately keep one sign. An infinite radius of
convergence with coefficients of one sign is not excluded for the Holstein
model, however, because the mass grows exponentially rather than as a power
of the coupling at strong coupling~\cite{LangFirsov1962}; the atomic limit is
an explicit example. Whether the radius is finite at $t>0$ is not settled by
four coefficients, although the growing ratios for $\omega_0\lesssim2t$
suggest that it is. The coefficients $d_n$ of the inverse mass already change
sign at low order when $\omega_0\gg t$ (at $\omega_0=8t$, $d_1$ and $d_2$ are
negative and $d_3$ and $d_4$ positive).

\emph{Borel-Pad\'e continuation.} The conclusion of Sec.~\ref{sec:borel} is
unchanged. At $\omega_0=t$ the Borel-Leroy $[3/1]$ approximant has a pole on
the positive axis for every Leroy parameter tried ($b=0,1,2,4$), and the
$[2/2]$ approximant for $b=1,2,4$. The pole-free $[2/2]$ approximant with
$b=0$ gives $m^*/m=1.38$ at $\lambda=1$, which is below the partial sum of
the series, $1.43$, all of whose terms are positive, and far below the exact
value $1.61$, whereas the ordinary $[2/2]$ Pad\'e approximant gives $1.54$.

In summary, the qualitative conclusions of this paper hold for both models at
the order to which the series are known: the weak-coupling series converges,
its ordinary Pad\'e continuation is quantitatively useful up to roughly $70\%$
of the crossover coupling, Borel methods are neither justified nor useful,
and the weak-coupling coefficients do not determine how the polaron passes
through the crossover. Quantitatively the Fr\"ohlich series is by far the
better behaved of the two. Its coefficients are regular enough to determine a
radius of convergence to about one percent, and that radius coincides with
the crossover. The Holstein coefficients are known exactly, but they are much
less regular; their apparent singularity is weak and is tied to the crossover
only when $\omega_0$ is comparable to $t$.

\bibliography{polaron_refs}

\end{document}